\documentclass[twocolumn]{aastex701} 
\usepackage[T1]{fontenc}
\usepackage{ae,aecompl}
\usepackage{hyperref}
\usepackage{url}

\usepackage{longtable}
\usepackage{booktabs}
\usepackage{tabu}
\usepackage{graphicx}
\usepackage{multirow}
\usepackage{ulem}
\usepackage{color}
\usepackage{amsmath}

\hypersetup{citecolor=blue, 
            linkcolor=red, 
            menucolor=blue, 
            urlcolor=blue}  

\def \kev{~\rm{keV}}

\def \cm{~\rm{cm}}
\def \s{~\rm{s}}
\def \km{~\rm{km}}

\def \K{~\rm{K}}

\def \g{~\rm{g}}
\def \G{~\rm{G}}

\def \erg{~\rm{erg}}

\def \keV{~\rm{keV}}

\usepackage{xcolor}
\definecolor{redak}{rgb}{0.9,0.15,0.05}

\shorttitle{Point symmetry in SNR G11.2‑0.3}
\shortauthors{Soker}

\graphicspath{{./}{figures/}}

\begin{document}

\title{Point-symmetric morphology in supernova remnant G11.2-0.3: the jittering jets explosion mechanism}

\author[0000-0003-0375-8987]{Noam Soker} 
\affiliation{Department of Physics, Technion Israel Institute of Technology, Haifa, 3200003, Israel; soker@physics.technion.ac.il}
\email{soker@physics.technion.ac.il}

\date{\today}

\begin{abstract}
I identify a point-symmetric morphology in the core-collapse supernova (CCSN) remnant SNR G11.2-0.3 composed of three pairs of opposite morphological features, and attribute their shaping to three energetic pairs of jets during the explosion process in the frame of the jittering jets explosion mechanism (JJEM). The pairs of morphological features are two opposite rings, a strip of dense ejecta extending on both sides of the central pulsar PSR J1811–1925, and an ear-nozzle opposite structure. According to the JJEM, additional weaker pairs of jets may also have participated in the explosion. The jets’ axis from the ear to the nozzle coincides with the axis of the presently active pulsar jets, which is the pulsar spin axis. The jets of this pair were the last that the newly born neutron star launched during the explosion, and the accretion disk that launched these jets spun up the neutron star in the same direction as the jets. The identification of a point-symmetric morphology in SNR G11.2-0.3 strengthens the claim that the JJEM is the primary explosion mechanism of CCSNe. 
\end{abstract}

\keywords{Supernova remnants -- Massive stars	--  Circumstellar material -- Stellar jets -- Supernova: individual (SNR G11.2‑0.3)}



\section{Introduction} 
\label{sec:intro}

The delayed neutrino mechanism (e.g., \citealt{Bambaetal2025CasA, Bocciolietal2025, BoccioliRoberti2025, EggenbergerAndersenetal2025, FangQetal2025, Huangetal2025, Imashevaetal2025, Janka2025, Maltsevetal2025, Maunderetal2025, Morietal2025, Mulleretal2025, Nakamuraetal2025, SykesMuller2025, Janka2025, Orlandoetal20251987A, ParadisoCoughlin2025, PowellMuller2025, Tsunaetal2025, Vinketal2025, WangBurrows2025, Willcoxetal2025, Mukazhanov2025, Raffeltetal2025, Vartanyanetal2025, Giudicietal2026} for papers since 2025),  and the jittering jets explosion mechanism (JJEM; e.g., \citealt{Bearetal2025Puppis, BearSoker2025, Braudoetal2025, Shishkinetal2025S147, Soker2025G0901, Soker2025N132D, Soker2025RCW89, Soker2025Learning, Soker2025Dust, SokerAkashi2025, SokerShiran2025, SokerShishkin2025Vela, SokerShishkin2025W49B, WangShishkinSoker2025} for papers since 2025) are the most intensively studied competing theoretical explosion mechanisms of core-collapse supernovae (CCSNe). Each of these two mechanisms aims to explain most CCSNe as the primary explosion mechanism. Other energy sources might provide additional energy to the exploding massive star, while other mechanisms have been developed to explain a minority of CCSNe. However, only the neutrino-driven mechanism\footnote{See  \citealt{Janka2025Padova} for a recent talk on the neutrino-driven mechanism: \url{https://www.memsait.it/videomemorie/volume-2-2025/VIDEOMEM_2_2025.46.mp4}} and the JJEM\footnote{See \citealt{Soker2025Padova} for a talk on the JJEM: \url{https://www.memsait.it/videomemorie/volume-2-2025/VIDEOMEM_2_2025.47.mp4}} have been heavily studied and presented in meetings in recent years as the primary explosion mechanisms of CCSNe. 

Some recent studies (e.g., \citealt{Maltsevetal2025}) confuse the role of the JJEM as a primary explosion mechanism with additional energy sources and rare theoretical explosion mechanisms (most other studies of the neutrino-driven mechanism ignore the JJEM altogether). Therefore, I present these mechanisms and gravitational energy sources in Table \ref{Tab:Table} to clarify their relationships. Explosion by thermonuclear burning triggered by the collapse of a pre-collapse mixed layer of helium and oxygen (e.g., \citealt{KushnirKatz2015, BlumKushnir2016}) might at best explain a small fraction of CCSNe because the helium-oxygen mixed layer with the required rotation is rare (e.g., \citealt{Gofmanetal2018}). Indeed, this model has not been studied in recent years. So, I do not include it in the table. 
\begin{table*}   
\begin{center}
\caption{Energy sources in the neutrino-driven mechanism and the JJEM}
\label{Tab:Table}
\begin{tabular}{|c|c|c|c|}
\hline
\textbf{Energy carrier}    & \multicolumn{2}{|c|}{\underline{ \textbf{Primary explosion mechanism:}}}   & \textbf{Comments}  \\
\textbf{(Carrier source)}    & \textbf{Neutrino-driven}   & \textbf{JJEM}   &   \\
\hline
Neutrino heating; & The primary & Additional energy$^{[1]}$  & Similar neutrino emission \\ 
(NS cooling)      & explosion process  &    &  in both mechanisms \\   
\hline
 Pairs of jittering jets;   & Jets do not exist  & The primary    & Point-symmetric CCSNRs  \\
(Stochastic $\vec{J}$ accretion) &   & explosion process   &  support the JJEM  \\
\hline
 Magnetorotational jets & Operates in rare energetic & The extreme-rare end of the    &     \\
 (rapidly-rotating core) & CCSNe; leaves NSs  & JJEM; not a separate process &    \\
\hline
 Collapsar jets   & Neutrino-mechanism fails; &  Part of the JJEM$^{[2]}$; &  \\ 
 (rapidly-rotating core)   & rare CCSNe that form BHs  &  rare CCSNe  that form BHs  &  \\
 \hline
 Magnetar  &  Common extra energy   & Possible extra energy source   & Most models with magnetars     \\ 
 (rapidly-rotating NS + $\vec{B}$) & in super-energetic CCSNe &    &  require explosion by jets$^{[3]}$   \\
  \hline
   Post-explosion accretion & Possible extra energy source  & Possible extra energy source   & \\ 
    (fallback material)    &                               & of late jet's pairs      &\\
\hline
\end{tabular}
\end{center}
The relationship between two theoretical mechanisms, each separately aiming to explain most CCSNe, and gravitational energy sources. 
Notes:
$\vec{J}$: angular momentum of accreted gas, which varies stochastically in magnitude and direction; $\vec B$: magnetic field of the newly born NS; BH: black hole; NS: neutron star; JJEM: jittering jets explosion mechanism; CCSNR: core-collapse supernova remnant. 
References: 
[1] \cite{Soker2022nu}; 
[2] \cite{Soker2023gap};  
[3] \cite{SokerGilkis2017, Kumar2025}; 
\end{table*}

The table presents two theoretical explosion mechanisms that aim to explain most CCSNe: the neutrino-driven (delayed neutrino explosion) mechanism, which revives the stalled shock by heating the post-shock material with neutrinos emitted by the cooling neutron star (NS), and the JJEM. In the JJEM, accretion of gas with stochastic angular momentum magnitude and direction onto the newly born NS launches the jittering jets that explode the star. The jets in most cases are not relativistic, i.e., their velocity is $<0.5c$; some studies support this claim (e.g., \citealt{Izzoetal2019} suggested jets at $\simeq 10^5 \km \s^{-1}$ in SN 2017iuk associated with GRB 171205A, and \citealt{Guettaetal2020} claimed that most CCSNe have no signatures of relativistic jets). 
In the JJEM, neutrino heating boosts the explosion (\citealt{Soker2022nu}), but the launching of jittering jets is the primary explosion process. Even in cases where neutrino heating might have revived the shock and driven a CCSN explosion, jittering jets are likely to operate earlier and explode the star in the frame of the JJEM \citep{WangShishkinSoker2025}.

The magnetorotational explosion mechanism can at best explain only a small fraction of CCSNe, as it requires the pre-collapse core to have fast rotation (e.g., \citealt{Shankaretal2025}); the explosion is driven by a single pair of long-lasting jets along a single axis (e.g., \citealt{Shibataetal2025}). Studies of the magnetorotational explosion mechanism assert that the neutrino-driven mechanism explodes most CCSNe (e.g., \citealt{Shankaretal2025}). According to the JJEM, the magnetorotational explosion mechanism is not a separate mechanism, but rather the extreme edge of the JJEM where the specific angular momentum of the pre-collapse core is much larger than the amplitudes of the stochastic specific angular momentum component of the accreted gas. The jets jitter at small angles around the fixed axis. 
If a black hole forms at the center and launches jets, the mechanism is a collapsar (e.g., \citealt{BoppGottlieb2025, Gottliebetal2025}, for recent papers); this is also a rare event. Another process that can cause stochastic changes in the jets' directions is wobbling, resulting from instability due to jet-disk interaction, as simulations have shown for black holes, namely collapsars (e.g., \citealt{Gottliebetal2022, Gottlieb2025}). 
 
In the JJEM, black holes form in a fraction of cases with pre-collapse, rapidly rotating cores. The jets jitter close to a fixed axis, and do not expel mass from around the equatorial plane; the accretion of this mass forms the black hole (e.g., \citealt{Gilkisetal2016, Soker2023gap}). This might lead to an energetic explosion; in the JJEM, there are no `failed supernovae' even during black hole formation. The neutrino-driven mechanism predicts that a non-negligible fraction of massive stars end in failed supernovae, leading to low-energy transients rather than CCSN explosions (e.g., \citealt{Antonietal2025} for a recent study).

The magnetar (e.g., \citealt{Blanchardetal2026}) and post-explosion accretion are not explosion processes, as they require an explosion to add energy to the ejecta. Many models of energetic CCSNe with magnetars find explosion energies of $E_{\rm exp} \gtrsim 3 \times 10^{51} \erg$ (e.g., \citealt{Aguilaretal2025, Orellanaetal2025}, as recent examples) that imply explosion by jets (e.g., \citealt{SokerGilkis2017, Kumar2025}). Magnetars, therefore, cannot save the neutrino-driven mechanism from its energy crisis, i.e., explaining  CCSNe with $E_{\rm exp} \gtrsim 2 \times 10^{51} \erg$. Post-explosion accretion can add energy. In the JJEM, this is a natural continuation of the JJEM itself as the fallback material forms accretion disks that launch jets. 

Many observables, like neutrino spectrum, are similar to the neutrino-driven mechanism and the JJEM, leaving the morphologies of CCSN remnants (CCSNRs) as the only observable that decisively distinguishes between the two mechanisms (\citealt{Soker2024UnivReview, Soker2025Learning} for reviews). 
Generally, supernova remnants (SNRs) can reveal much information about the physics of supernovae and the processes involved in the interaction of their ejecta with the surrounding medium (e.g., \citealt{Yanetal2020RAA, Luetal2021RAA}), including jets (e.g., \citealt{YuFang2018RAA}), cosmic ray acceleration, emission properties (e.g., \citealt{Yamazakietal2014RAA, Zhangetal2016RAA, Lietal2020RAA, Luoetal2024RAA}), magnetohydrodynamics (e.g., \citealt{Wuetal2019RAA, Leietal2024RAA}), the role of the NS remnant  (e.g., \citealt{HorvathAllen2011RAA, Wuetal2021RAA}), and can reveal the structure of the ejecta (e.g., \citealt{Renetal2018RAA}).
The identification of 16 CCSN remnants (CCSNRs) with point-symmetric morphologies attributed to two or more pairs of jets (see list in \citealt{WangShishkinSoker2025}) strongly supports the JJEM as the primary explosion mechanism of CCSNe.  

The presence of shells in CCSNe might be an emerging observable to distinguish between the two explosion mechanisms. Simulations show that the launching of jittering jets compresses several shells in the ejecta, either full, i.e., covering a solid angle of $4 \pi$, or partial shells as caps of ears, lobes, or bubbles (e.g., \citealt{Braudoetal2025}). Indeed, several CCSNRs exhibit two or more shells. \cite{SokerShiran2025} analyzed the temporal evolution of the photosphere of the ejecta from SN 2023ixf, as calculated by \cite{Zimmermanetal2024}.  \cite{SokerShiran2025} concluded that there are three shells in the ejecta that contribute to the photosphere, and that this is compatible with the JJEM. In a recent study, \cite{YangYietal2025} concluded from their early spectropolarimetry of SN 2024ggi that it had a persisting, prominent symmetry axis throughout the explosion. They further propose a bipolar explosion. A bipolar explosion is compatible with a jet-driven explosion. 
\cite{DeSotoetal2026} find the polarization of SN 2012au to maintain a near-constant orientation during the early photospheric phase, then changing direction in the transition to the nebular phase. This asymmetry evolution, together with the explosion energy of SN 2012au of $\simeq 10^{52} \erg$ \citep{Milisavljevicetal2013}, which is much above what the neutrino-driven mechanism can yield, strongly indicates the JJEM. 

In this study, I identify a point-symmetric morphology in SNR G11.2-0.3. Papers have addressed several aspects of SNR G11.2-0.3 and its pulsar PSR J1811–1925 with its pulsar wind nebula (PWN; e.g., \citealt{Downes1984, Kaspietal2001, Kooetal2007, Moonetal2009, Leeetal2013, Borkowskietal2016, Chawneretal2019, Guestetal2020, Hiraietal2020, ZhangYetal2025}). I focus on the morphology. 
In Section \ref{sec:Rings} I identify a pair of rings that I attribute to a pair of jets that shaped them during the explosion. In section \ref{sec:InnerAxes} I identify two additional jets' axes that are prominent in the inner ejecta. I summarize this study in  Section \ref{sec:Summary} by further supporting the JJEM as the primary explosion mechanism of CCSNe.

\section{A pair of circum-jet rings}
\label{sec:Rings}

\begin{figure}[]
	\begin{center}
\includegraphics[trim=2.0cm 1.0cm 0.0cm 1.6cm ,clip, scale=0.66]{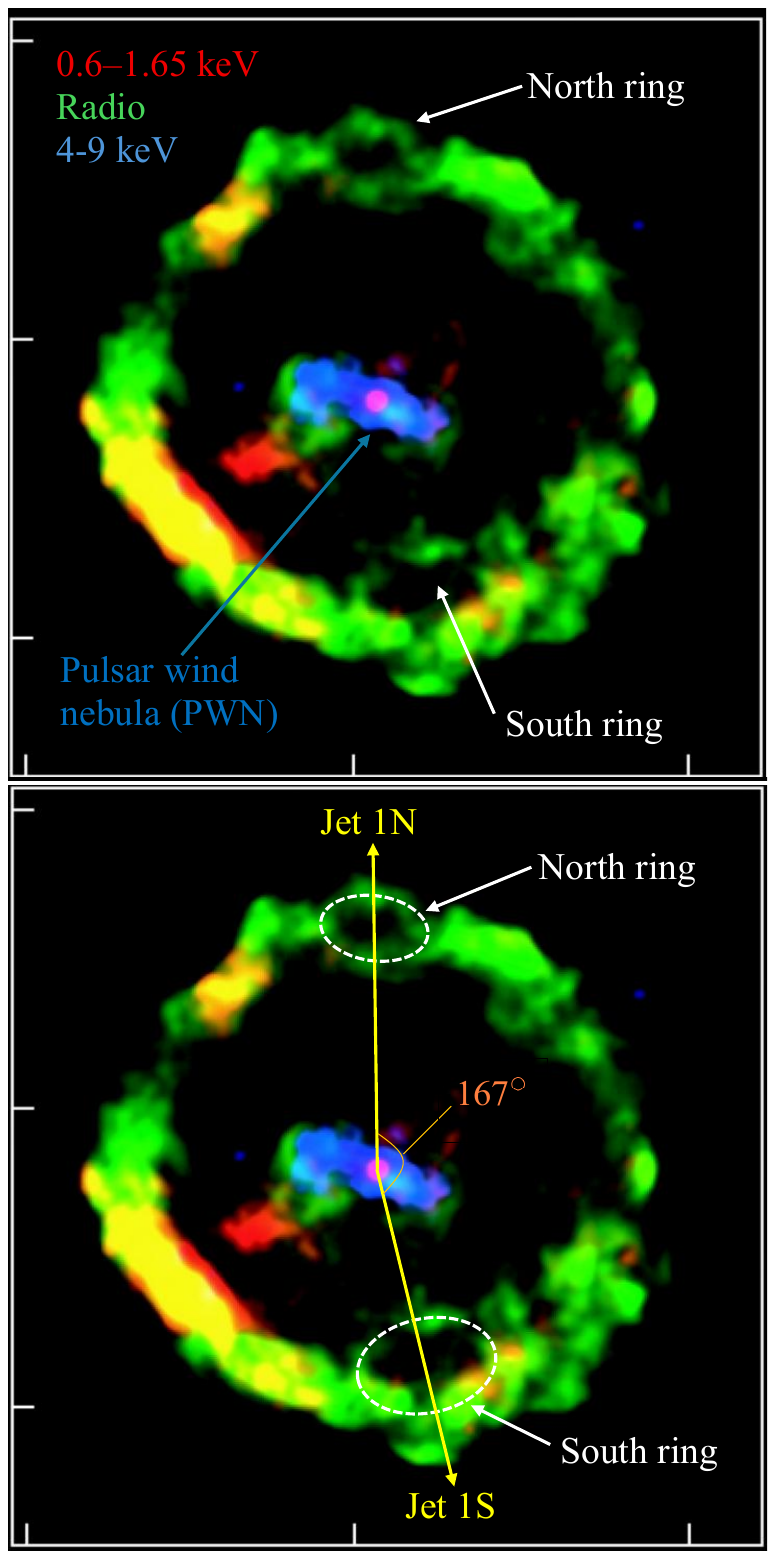} 
\caption{
A figure adapted from \cite{Robertsetal2003}, comparing X-ray pulsar wind nebula (PWN) emission and radio emission. Red: $0.6-1.65 \kev$ X-ray. Green: 3.5 cm radio. Blue: $4-9 \keV$ X-ray. All are at $5^{\prime \prime}$ resolution.
I added the identification of two rings and the directions of the two jets that I suggest shaped these circum-jet rings during the explosion. The two images are identical, allowing a clear view without marks in the upper panel. 
The two rings define Jet Pair 1. 
Right ascension (J2000) ticks are 18:11:40, 18:11:30 and 18:11:20, and declination (J2000) ticks are $-19:27:00$, $-19:25:00$, and $-19:23:00$. 
}
\label{Fig:G11FigureOuter1}
\end{center}
\end{figure}

Figure \ref{Fig:G11FigureOuter1} presents an image of SNR G11.2-0.3, adapted from \cite{Robertsetal2003}, which presents X-ray (red and blue) and radio (green) emission. The upper panel of Figure \ref{Fig:G11FigureOuter1} provides a clear view, with only a few marks that I added. The lower panel of the same image includes my identification of two opposite rings. I suggest that two jets inflated them during the explosion process, as in the simulations of \cite{SokerAkashi2025} for CCSNe and \cite{Akashietal2025} for planetary nebulae. Drawing from the present location of the pulsar, the two jets are bent by $13^\circ$ from being exactly opposite. This bent asymmetry is observed in other CCSNRs, e.g., one pair of bays in the Crab Nebula \citep{ShishkinSoker2025Crab}. 

Earlier studies suggested the formation of circum-jet rings in CCSNe by exploding jets, i.e., jets that participated in the explosion. These CCSNRs include  SNR 0540-69.3 \citep{Soker2022SNR0540}, W49B \citep{SokerShishkin2025W49B}, and four that \cite{SokerAkashi2025} discuss and compare to their simulations: the Cygnus Loop, SNR G0.9+0.1, SNR G107.7-5.1, and SNR Circinus X-1; \cite {Gasealahweetal2025} attributed the rings of SNR Circinus X-1 to post-explosion jets, rather than exploding jets. 
I emphasize that the jets that shaped the rings were active only during the explosion process (or shortly after), for about a second or less. The jets penetrated through a shell and compressed the shell material to the sides to form the rings \citep{SokerAkashi2025}. 

I fit the ellipses on the lower panel of Figure \ref{Fig:G11FigureOuter1} by eye. This is adequate for the present study. The axis ratios for the north and south ellipses are $0.6$ and $0.67$. Assuming that the rings are circular implies an inclination angle (jet axis to line of sight) of $i_{\rm 1N} \simeq 53^\circ$ and $i_{\rm 1S} \simeq 48^\circ$, respectively. The similar inclinations of the two opposite rings strengthen their identification as circum-jet rings of two opposite jets. I term this north-south pair of jets Pair 1.  

\section{Two symmetry axes by the inner region}
\label{sec:InnerAxes}

\begin{figure}[]
	\begin{center}
\includegraphics[trim=0.2cm 17.3cm 0.0cm 2.4cm ,clip, scale=0.64]{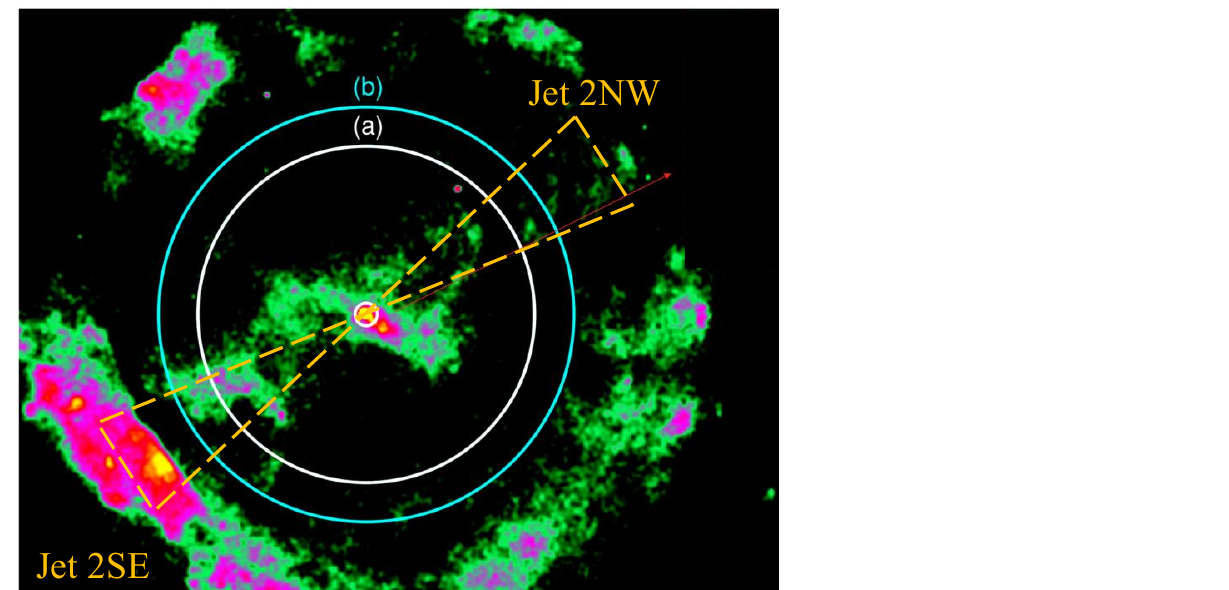} 
\caption{
A Chandra image adapted from \cite{ZhengJTetal2023RAA}; the two circles with their labels and the red arrow are their marking and irrelevant to this study. There is a bright inner strip extending from southeast to northwest. I suggest a shaping by a pair of jets, Pair 2. I mark the general directions of the two jets with the orange biconical, which represents the uncertainty in their directions rather than their shapes.     }
\label{Fig:G11Inner2}
\end{center}
\end{figure}

\begin{figure}[]
	\begin{center}
\includegraphics[trim=0.0cm 0.5cm 0.0cm 0.0cm ,clip, scale=0.55]{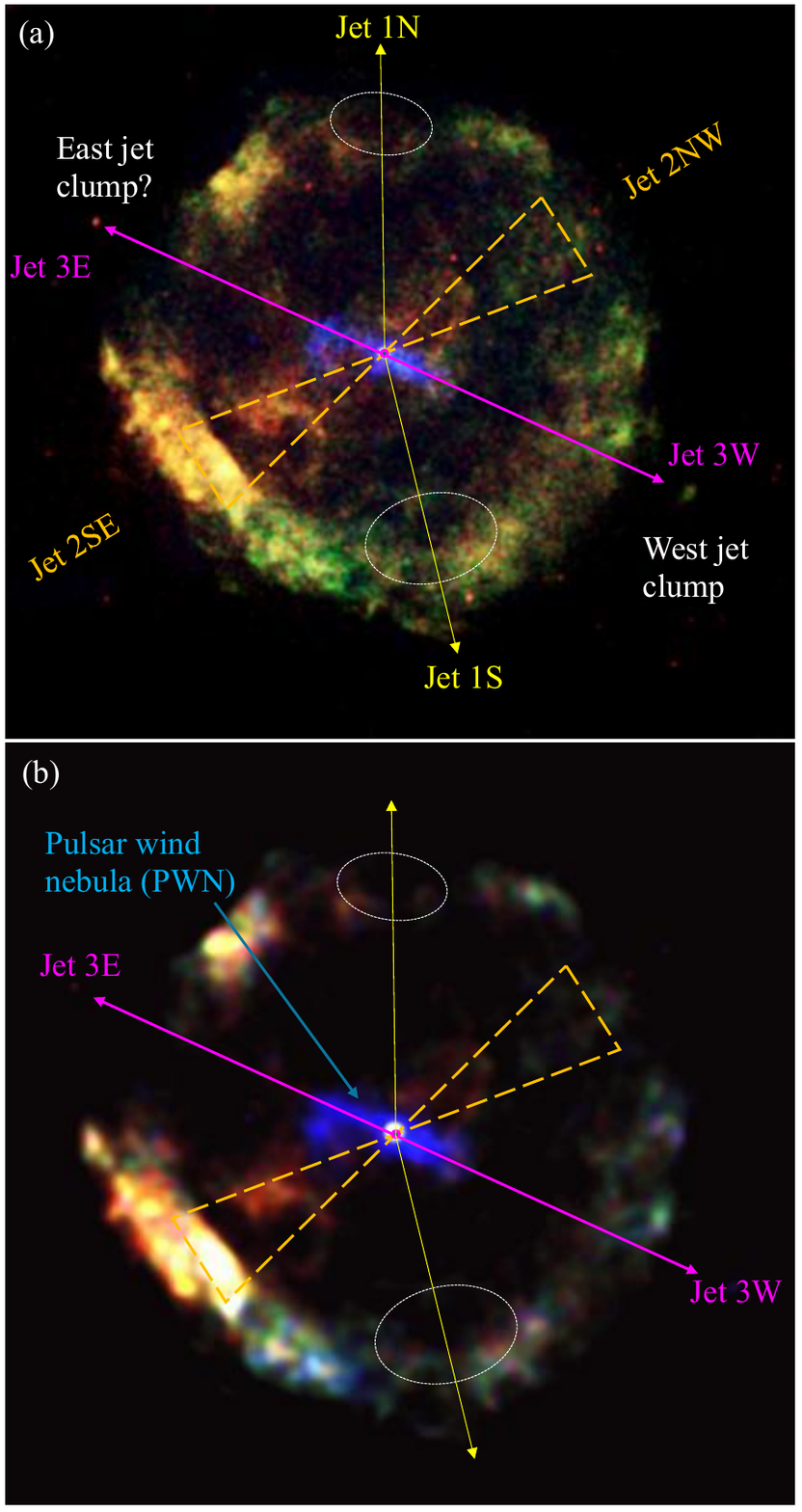} 
\caption{ Two images of SNR G11.2-0.3, with the three pairs of jets identified in this study. The ellipses and arrows of Pair 1 are as in Figure \ref{Fig:G11FigureOuter1} (and on the same relative scale), and the biconical shape is as in Figure \ref{Fig:G11Inner2}. The double-sided arrow defines Pair 3, two opposite jets that shaped the ear-nozzle structure during the explosion; its length has no meaning, as it only marks the jets' axis. The direction is as the present pulsar jets, but the jets of Pair 3 were active for only a second or so during the explosion. (a) An image of SNR G11.2-0.3 adapted from the \href{https://chandra.harvard.edu/photo/2007/g11/}{Chandra site}. Color code is for energy bands: red $0.5-1.5 \keV$; green $1.5-2.5 \keV$; blue $2.5-8 \keV$. 
(Credit: NASA/CXC/Eureka Scientific/\citealt{Robertsetal2003})
(b) Another image of SNR G11.2-0.3 from \href{https://chandra.harvard.edu/photo/2001/1227/}{Chandra site}. Color code is for energy bands: red $0.6-1.65 \keV$; green $1.65-2.25 \keV$; blue $2.25-7.5 \keV$. The intensity color codes of the two panels differ, and so do the scales; in panel (b), the SNR is $1.08$ as large as in panel (a). 
(Credit: NASA/McGill/\citealt{Kaspietal2001})
}
\label{Fig:G11Inner1}
\end{center}
\end{figure}

Figure \ref{Fig:G11Inner2} presents a Chandra X-ray image of SNR G11.2-0.3 adapted from \cite{ZhengJTetal2023RAA}. The bright central point is the pulsar. The bright horizontal bar near the center is the PWN, seen in blue (hard X-ray) in Figures \ref{Fig:G11FigureOuter1} and \ref{Fig:G11Inner1}. There is a bright strip extending from the southeast part of the main shell to the northwest, as seen in Figure \ref{Fig:G11Inner2} in green, and in Figures \ref{Fig:G11FigureOuter1} and \ref{Fig:G11Inner1} mainly in red (soft X-ray). This general structure is another pair of opposite morphological features, part of the point-symmetric structure of SNR G11.2-0.3. I suggest that an energetic pair of jets (Pair 2) shaped this strip; the jets were active only during the explosion. However, I cannot determine the exact direction of the jets, i.e., the pair's axis, and therefore I draw a biconical shape. The biconical shape is not the shape of the jets, but rather signifies the uncertainty in the axis of Pair 2.   

\begin{figure}[]
	\begin{center}
\includegraphics[trim=0.2cm 1.9cm 0.0cm 0.0cm ,clip, scale=0.54]{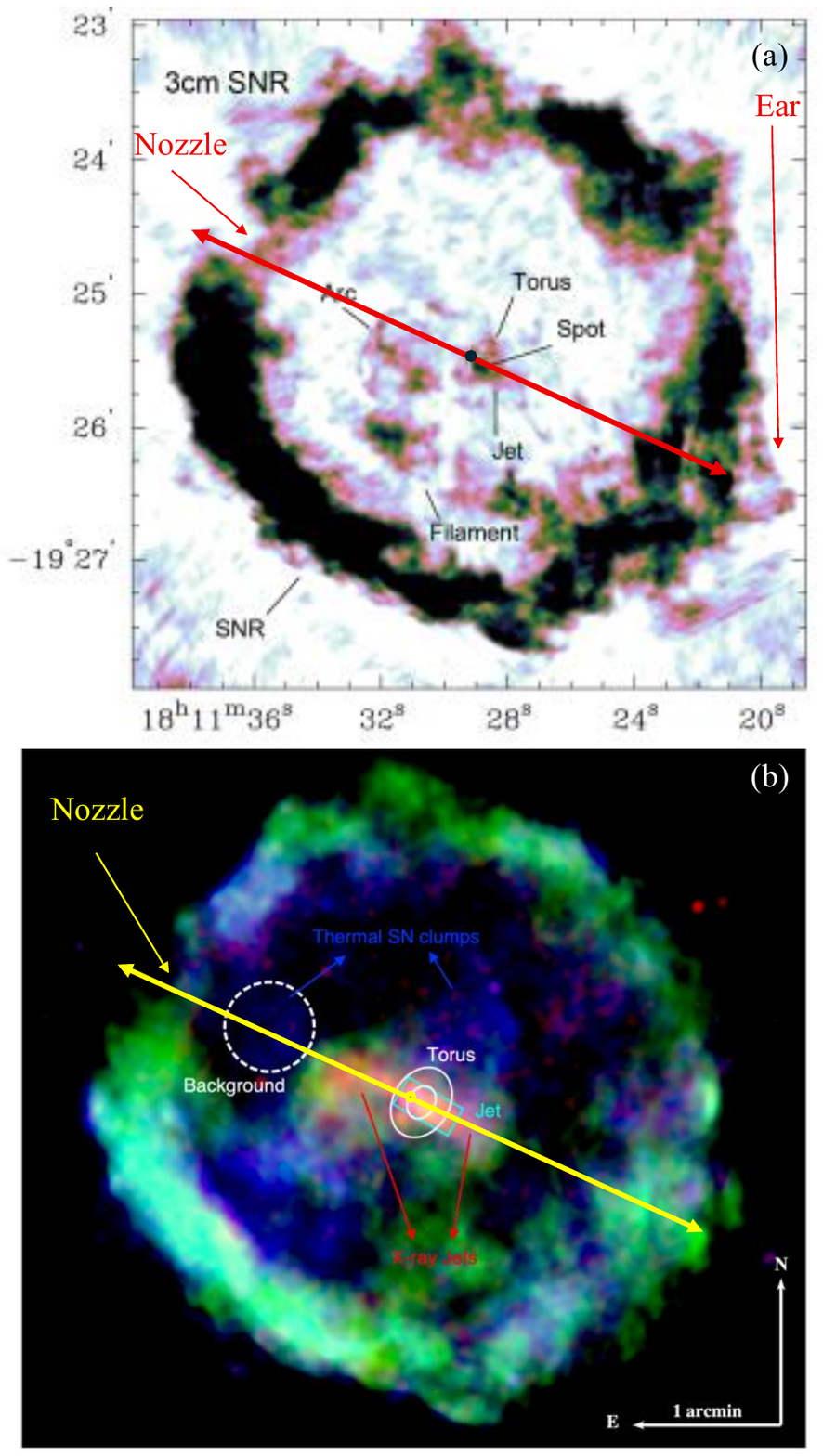}
\caption{ 
Two images of SNR G11.2-0.3 adapted from \cite{ZhangYetal2025}. I added a double-sided arrow through the pulsar at the center and the marks of the nozzle and ear. The double-sided arrow length is $4.4^{\prime}$, as in Figure \ref{Fig:G11Inner1}. 
(a) Australia Telescope Compact Array (ATCA) $3 \cm$ radio image. 
(b) RGB image: green: 6 cm ATCA radio map;  red: $2.7-9.0 \keV$ (Chandra); Blue: $0.5-2.0 \keV$ (Chandra). 
 }
\label{Fig:ZhangYetal2025}
\end{center}
\end{figure}

The jets of the pulsar are documented in the literature (e.g., \citealt{Kaspietal2001, Deanetal2008, Madsenetal2020} ). These jets of the pulsar do not seem to expand beyond the PWN. However, I identify morphological features attributed to jets in other CCSNRs in the outer regions of SNR G11.2-0.3.  In Figure \ref{Fig:ZhangYetal2025}, adapted from \cite{ZhangYetal2025}, I marked a nozzle and an ear. A nozzle is an opening in the main SNR shell. The different figures here show that the east-east-north segment is the faintest and largest opening in the bright shell, both in radio and X-ray emissions. On the opposite side, there is an ear, defined as a protrusion smaller than the main shell and with a decreasing cross-section with distance. Several astrophysical objects, such as planetary nebulae, exhibit this nozzle-rim asymmetry. In \cite{Soker2024PNSN}, I compared rim-nozzle asymmetry in planetary nebulae to four CCSNRs: SN 1987, G107.7-5.1, SNR G309.2–00.6, and the Vela SNR. On the nozzle side, the jet breaks out, leaving a nozzle. On the rim side, the jet does not break out, but rather compresses a cap or an ear.  

I attribute the opposite nozzle-ear structure of SNR G11.2-0.3 to shaping by a pair of jets (Pair 3).
The double-sided arrow that crosses through the center in Figures \ref{Fig:G11Inner1} and \ref{Fig:ZhangYetal2025} presents the direction of the jet axis I propose for Pair 3. The length of the double-sided arrow is the same in all figures, $4.4^{\prime}$. In panel (b) of Figure \ref{Fig:G11Inner1}, the tip of the ear appears as a clump, the west jet clump. On the opposite side, there is a bright point; I am not sure about its nature, but I encourage further observations to reveal its nature: is it a star unrelated to the SNR, or is it a clump associated with Pair 3?  

The jets of Pair 3 were in the same direction as the jets of the pulsar, presumably along the spin axis of the pulsar. According to the JJEM, several pairs of jets launched by the newly born NS explode the star. The newly born NS launches jets when it accretes gas with high specific angular momentum; the jets are launched along the angular momentum axis of the accreted gas. I suggest that Pair 3 was an energetic pair of jets and the last to occur during the explosion. The launching of the energetic jets resulted from the accretion of mass with angular momentum in the same direction as the jets, leaving the NS with a spin axis aligned with the jets. This explains why the pulsar's jets lie along Pair 3.

\section{Summary} 
\label{sec:Summary}

I identified a point-symmetric morphology in SNR G11.2-0.3 composed of three pairs of opposite morphological features. I attribute these to three pairs of energetic jets that participated in the explosion process in the framework of the JJEM. The jets that shaped the point-symmetric morphology were active for several seconds or less during the explosion. According to the JJEM, additional weaker pairs of jets may also have participated in the explosion. 

Figure \ref{Fig:G11FigureOuter1} presents two opposite rings on the main SNR shell. The arrows present the jets' direction. The two jets that shaped them are at $167^\circ$ (projected on the plane of the sky) to each other with respect to the present location of the pulsar, but they belong to the same jet-launching episode during the explosion. This is Pair 1, with the jets' axes inclined at $\simeq 50^\circ$ to the line of sight. Figure \ref{Fig:G11Inner2} presents a bright strip, which appears in red in Figure \ref{Fig:G11FigureOuter1} and \ref{Fig:G11Inner1}, extending to both sides of the pulsar, from the southeast side of the main SNR shell to its northwest side. The orange biconical signifies the uncertainty in the axis of the jets that compose Pair 2, rather than the shape of the jets. In Figure  \ref{Fig:ZhangYetal2025}, I mark the ear-nozzle opposite pair, which I attribute to Pair 3 of exploding jets. Figure \ref{Fig:G11Inner1} presents the three pairs of jets as indicated by arrows and a biconical shape, which are the estimated directions of the jets. 

The axis of Pair 3 is the same as that of the still active pulsar jets. However, the ear-nozzle structure requires more energetic jets than those of the pulsar. According to the JJEM, Pair 3 was the last one launched by the newly born NS. The accreted gas during this jet-launching episode spun up the NS to have a spin in the same direction as the jets, i.e., the angular momentum axis of the accreted gas. The present pulsar, therefore, has its spin in the same direction. 
    
The JJEM predicts that many CCSNRs should have point-symmetric morphologies. Because several processes smear the point-symmetric morphologies (see \citealt{SokerShishkin2025Vela} for a discussion), not all CCSNRs exhibit point-symmetric morphologies that are prominent enough for detection. There are about 20 CCSNRs with identified point-symmetric morphologies attributed to the JJEM. The competing neutrino-driven mechanism cannot explain these observations. As such, the identification of a point-symmetric morphology in SNR G11.2-0.3 strengthens the claim that the JJEM is the primary explosion mechanism of CCSNe.

\section*{Acknowledgements}
 
I thank Ealeal Bear, Dmitry Shishkin, and Aleksei Klimov for helpful discussions. The Charles Wolfson Academic Chair at the Technion supported this research.


%
\bibliography{reference}{}

@ARTICLE{Gilkisetal2016,
       author = {{Gilkis}, Avishai and {Soker}, Noam and {Papish}, Oded},
        title = "{Explaining the Most Energetic Supernovae with an Inefficient Jet-feedback Mechanism}",
      journal = {\apj},
         year = 2016,
        month = aug,
       volume = {826},
       number = {2},
          eid = {178},
        pages = {178},
          doi = {10.3847/0004-637X/826/2/178},
archivePrefix = {arXiv},
       eprint = {1511.01471},
 primaryClass = {astro-ph.HE},
       adsurl = {https://ui.adsabs.harvard.edu/abs/2016ApJ...826..178G}
}

@ARTICLE{Gottliebetal2022,
       author = {{Gottlieb}, Ore and {Liska}, Matthew and {Tchekhovskoy}, Alexander and {Bromberg}, Omer and {Lalakos}, Aretaios and {Giannios}, Dimitrios and {M{\"o}sta}, Philipp},
        title = "{Black Hole to Photosphere: 3D GRMHD Simulations of Collapsars Reveal Wobbling and Hybrid Composition Jets}",
      journal = {\apjl},
         year = 2022,
        month = jul,
       volume = {933},
       number = {1},
          eid = {L9},
        pages = {L9},
          doi = {10.3847/2041-8213/ac7530},
archivePrefix = {arXiv},
       eprint = {2204.12501},
 primaryClass = {astro-ph.HE},
       adsurl = {https://ui.adsabs.harvard.edu/abs/2022ApJ...933L...9G}
}

@ARTICLE{Guettaetal2020,
       author = {{Guetta}, Dafne and {Rahin}, Roi and {Bartos}, Imre and {Della Valle}, Massimo},
        title = "{Constraining the fraction of core-collapse supernovae harbouring choked jets with high-energy neutrinos}",
      journal = {\mnras},
         year = 2020,
        month = feb,
       volume = {492},
       number = {1},
        pages = {843-847},
          doi = {10.1093/mnras/stz3245},
archivePrefix = {arXiv},
       eprint = {1906.07399},
 primaryClass = {astro-ph.HE},
       adsurl = {https://ui.adsabs.harvard.edu/abs/2020MNRAS.492..843G}
}

@ARTICLE{Soker2022SNR0540,
       author = {{Soker}, Noam},
        title = "{Imprints of the Jittering Jets Explosion Mechanism in the Morphology of the Supernova Remnant SNR 0540-69.3}",
      journal = {Research in Astronomy and Astrophysics},
         year = 2022,
        month = mar,
       volume = {22},
       number = {3},
          eid = {035019},
        pages = {035019},
          doi = {10.1088/1674-4527/ac49e6},
archivePrefix = {arXiv},
       eprint = {2109.10230},
 primaryClass = {astro-ph.HE},
       adsurl = {https://ui.adsabs.harvard.edu/abs/2022RAA....22c5019S}
}

@ARTICLE{Soker2023gap,
       author = {{Soker}, Noam},
        title = "{The Neutron Star to Black Hole Mass Gap in the Frame of the Jittering Jets Explosion Mechanism (JJEM)}",
      journal = {Research in Astronomy and Astrophysics},
         year = 2023,
        month = sep,
       volume = {23},
       number = {9},
          eid = {095020},
        pages = {095020},
          doi = {10.1088/1674-4527/ace9b3},
archivePrefix = {arXiv},
       eprint = {2304.05705},
 primaryClass = {astro-ph.HE},
       adsurl = {https://ui.adsabs.harvard.edu/abs/2023RAA....23i5020S}
}

@ARTICLE{Soker2024PNSN,
       author = {{Soker}, Noam},
        title = "{Planetary Nebula Morphologies Indicate a Jet-Driven Explosion of SN 1987A and Other Core-Collapse Supernovae}",
      journal = {Galaxies},
         year = 2024,
        month = jun,
       volume = {12},
       number = {3},
          eid = {29},
        pages = {29},
          doi = {10.3390/galaxies12030029},
archivePrefix = {arXiv},
       eprint = {2404.14843},
 primaryClass = {astro-ph.HE},
       adsurl = {https://ui.adsabs.harvard.edu/abs/2024Galax..12...29S}
}

@ARTICLE{Soker2022nu,
       author = {{Soker}, Noam},
        title = "{Boosting Jittering Jets by Neutrino Heating in Core Collapse Supernovae}",
      journal = {Research in Astronomy and Astrophysics},
         year = 2022,
        month = sep,
       volume = {22},
       number = {9},
          eid = {095007},
        pages = {095007},
          doi = {10.1088/1674-4527/ac7cbc},
archivePrefix = {arXiv},
       eprint = {2202.05556},
 primaryClass = {astro-ph.HE},
       adsurl = {https://ui.adsabs.harvard.edu/abs/2022RAA....22i5007S}
}

@ARTICLE{Nakamuraetal2025,
       author = {{Nakamura}, Ko and {Takiwaki}, Tomoya and {Matsumoto}, Jin and {Kotake}, Kei},
        title = "{Three-dimensional magnetohydrodynamic simulations of core-collapse supernovae - I. Hydrodynamic evolution and protoneutron star properties}",
      journal = {\mnras},
         year = 2025,
        month = jan,
       volume = {536},
       number = {1},
        pages = {280-294},
          doi = {10.1093/mnras/stae2611},
archivePrefix = {arXiv},
       eprint = {2405.08367},
 primaryClass = {astro-ph.HE},
       adsurl = {https://ui.adsabs.harvard.edu/abs/2025MNRAS.536..280N}
}

@ARTICLE{ShishkinSoker2025Crab,
       author = {{Shishkin}, Dmitry and {Soker}, Noam},
        title = "{Et tu, Brute?: The Crab Nebula also exploded by jittering jets}",
      journal = {arXiv e-prints},
         year = 2024,
        month = nov,
          eid = {arXiv:2411.07938},
        pages = {arXiv:2411.07938},
archivePrefix = {arXiv},
       eprint = {2411.07938},
 primaryClass = {astro-ph.HE},
       adsurl = {https://ui.adsabs.harvard.edu/abs/2024arXiv241107938S}
}

@ARTICLE{SokerGilkis2017,
       author = {{Soker}, Noam and {Gilkis}, Avishai},
        title = "{Magnetar-powered Superluminous Supernovae Must First Be Exploded by Jets}",
      journal = {\apj},
         year = 2017,
        month = dec,
       volume = {851},
       number = {2},
          eid = {95},
        pages = {95},
          doi = {10.3847/1538-4357/aa9c83},
archivePrefix = {arXiv},
       eprint = {1708.08356},
 primaryClass = {astro-ph.HE},
       adsurl = {https://ui.adsabs.harvard.edu/abs/2017ApJ...851...95S}
}

@ARTICLE{Kumar2025,
       author = {{Kumar}, Amit},
        title = "{Insights from modelling magnetar-driven light curves of stripped-envelope supernovae}",
      journal = {\na},
         year = 2025,
        month = may,
       volume = {116},
          eid = {102346},
        pages = {102346},
          doi = {10.1016/j.newast.2024.102346},
archivePrefix = {arXiv},
       eprint = {2412.09357},
 primaryClass = {astro-ph.HE},
       adsurl = {https://ui.adsabs.harvard.edu/abs/2025NewA..11602346K}
}

@ARTICLE{Soker2025Learning,
       author = {{Soker}, Noam},
        title = "{Learning from core-collapse supernova remnants on the explosion mechanism}",
      journal = {\na},
         year = 2025,
        month = dec,
       volume = {121},
          eid = {102453},
        pages = {102453},
          doi = {10.1016/j.newast.2025.102453},
archivePrefix = {arXiv},
       eprint = {2409.13657},
 primaryClass = {astro-ph.HE},
       adsurl = {https://ui.adsabs.harvard.edu/abs/2025NewA..12102453S}
}

@ARTICLE{Bocciolietal2025,
       author = {{Boccioli}, Luca and {Vartanyan}, David and {O'Connor}, Evan P. and {Kasen}, Daniel},
        title = "{Neutrino heating in 1D, 2D, and 3D core-collapse supernovae: characterizing the explosion of high-compactness stars}",
      journal = {\mnras},
         year = 2025,
        month = jul,
       volume = {540},
       number = {4},
        pages = {3885-3905},
          doi = {10.1093/mnras/staf963},
archivePrefix = {arXiv},
       eprint = {2501.06784},
 primaryClass = {astro-ph.HE},
       adsurl = {https://ui.adsabs.harvard.edu/abs/2025MNRAS.540.3885B}
}

@ARTICLE{Soker2024UnivReview,
       author = {{Soker}, Noam},
        title = "{The Two Alternative Explosion Mechanisms of Core-Collapse Supernovae: 2024 Status Report}",
      journal = {Universe},
         year = 2024,
        month = dec,
       volume = {10},
       number = {12},
          eid = {458},
        pages = {458},
          doi = {10.3390/universe10120458},
archivePrefix = {arXiv},
       eprint = {2411.08555},
 primaryClass = {astro-ph.HE},
       adsurl = {https://ui.adsabs.harvard.edu/abs/2024Univ...10..458S}
}

@ARTICLE{Bearetal2025Puppis,
       author = {{Bear}, Ealeal and {Shishkin}, Dmitry and {Soker}, Noam},
        title = "{The Puppis A Supernova Remnant: An Early Jet-driven Neutron Star Kick followed by Jittering Jets}",
      journal = {Research in Astronomy and Astrophysics},
         year = 2025,
        month = apr,
       volume = {25},
       number = {4},
          eid = {045008},
        pages = {045008},
          doi = {10.1088/1674-4527/adc24e},
archivePrefix = {arXiv},
       eprint = {2409.11453},
 primaryClass = {astro-ph.HE},
       adsurl = {https://ui.adsabs.harvard.edu/abs/2025RAA....25d5008B}
}

@ARTICLE{Shibataetal2025,
       author = {{Shibata}, Masaru and {Fujibayashi}, Sho and {Wanajo}, Shinya and {Ioka}, Kunihito and {Lam}, Alan Tsz-Lok and {Sekiguchi}, Yuichiro},
        title = "{Self-consistent scenario for jet and stellar explosions in collapsar: General relativistic magnetohydrodynamics simulation with a dynamo}",
      journal = {\prd},
         year = 2025,
        month = jun,
       volume = {111},
       number = {12},
          eid = {123017},
        pages = {123017},
          doi = {10.1103/msy2-fwhx},
archivePrefix = {arXiv},
       eprint = {2502.02077},
 primaryClass = {astro-ph.HE},
       adsurl = {https://ui.adsabs.harvard.edu/abs/2025PhRvD.111l3017S}
}

@ARTICLE{Imashevaetal2025,
       author = {{Imasheva}, Liliya and {Janka}, Hans-Thomas and {Weiss}, Achim},
        title = "{Comparison of three methods for triggering core-collapse supernova explosions in spherical symmetry}",
      journal = {\mnras},
         year = 2025,
        month = jul,
       volume = {541},
       number = {1},
        pages = {116-134},
          doi = {10.1093/mnras/staf865},
archivePrefix = {arXiv},
       eprint = {2501.13172},
 primaryClass = {astro-ph.HE},
       adsurl = {https://ui.adsabs.harvard.edu/abs/2025MNRAS.541..116I}
}

@ARTICLE{SokerShishkin2025Vela,
       author = {{Soker}, Noam and {Shishkin}, Dmitry},
        title = "{The Vela Supernova Remnant: The Unique Morphological Features of Jittering Jets}",
      journal = {Research in Astronomy and Astrophysics},
         year = 2025,
        month = mar,
       volume = {25},
       number = {3},
          eid = {035008},
        pages = {035008},
          doi = {10.1088/1674-4527/adb4cc},
archivePrefix = {arXiv},
       eprint = {2409.02626},
 primaryClass = {astro-ph.HE},
       adsurl = {https://ui.adsabs.harvard.edu/abs/2025RAA....25c5008S}
}

@ARTICLE{BearSoker2025,
       author = {{Bear}, Ealeal and {Soker}, Noam},
        title = "{Identifying a point-symmetric morphology in supernova remnant Cassiopeia A: Explosion by jittering jets}",
      journal = {\na},
         year = 2025,
        month = jan,
       volume = {114},
          eid = {102307},
        pages = {102307},
          doi = {10.1016/j.newast.2024.102307},
archivePrefix = {arXiv},
       eprint = {2403.07625},
 primaryClass = {astro-ph.HE},
       adsurl = {https://ui.adsabs.harvard.edu/abs/2025NewA..11402307B}
}

@ARTICLE{Soker2025G0901,
       author = {{Soker}, Noam},
        title = "{The Morphology of Supernova Remnant G0.9+0.1 Implies Explosion by Jittering-jets}",
      journal = {Research in Astronomy and Astrophysics},
         year = 2025,
        month = nov,
       volume = {25},
       number = {11},
          eid = {115005},
        pages = {115005},
          doi = {10.1088/1674-4527/adfd23},
archivePrefix = {arXiv},
       eprint = {2504.11384},
 primaryClass = {astro-ph.HE},
       adsurl = {https://ui.adsabs.harvard.edu/abs/2025RAA....25k5005S}
}

@ARTICLE{Huangetal2025,
       author = {{Huang}, Xu-Run and {Zha}, Shuai and {Chu}, Ming-chung and {O'Connor}, Evan P. and {Chen}, Lie-Wen},
        title = "{Phase-transition-induced Collapse of Proto-compact Stars and Its Implication for Supernova Explosions}",
      journal = {\apj},
         year = 2025,
        month = feb,
       volume = {979},
       number = {2},
          eid = {151},
        pages = {151},
          doi = {10.3847/1538-4357/ada146},
archivePrefix = {arXiv},
       eprint = {2409.16189},
 primaryClass = {astro-ph.HE},
       adsurl = {https://ui.adsabs.harvard.edu/abs/2025ApJ...979..151H}
}

@ARTICLE{EggenbergerAndersenetal2025,
       author = {{Eggenberger Andersen}, Oliver and {O'Connor}, Evan and {Andresen}, Haakon and {da Silva Schneider}, Andr{\'e} and {Couch}, Sean M.},
        title = "{Black Hole Supernovae, Their Equation of State Dependence, and Ejecta Composition}",
      journal = {\apj},
         year = 2025,
        month = feb,
       volume = {980},
       number = {1},
          eid = {53},
        pages = {53},
          doi = {10.3847/1538-4357/ada899},
archivePrefix = {arXiv},
       eprint = {2411.11969},
 primaryClass = {astro-ph.HE},
       adsurl = {https://ui.adsabs.harvard.edu/abs/2025ApJ...980...53E}
}

@ARTICLE{Maunderetal2025,
       author = {{Maunder}, Thomas and {Callan}, Fionntan P. and {Sim}, Stuart A. and {Heger}, Alexander and {M{\"u}ller}, Bernhard},
        title = "{Synthetic Light Curves and Spectra for the Photospheric Phase of a 3D Stripped-Envelope Supernova Explosion Model}",
      journal = {arXiv e-prints},
         year = 2024,
        month = oct,
          eid = {arXiv:2410.20829},
        pages = {arXiv:2410.20829},
          doi = {10.48550/arXiv.2410.20829},
archivePrefix = {arXiv},
       eprint = {2410.20829},
 primaryClass = {astro-ph.HE},
       adsurl = {https://ui.adsabs.harvard.edu/abs/2024arXiv241020829M}
}

@ARTICLE{Mulleretal2025,
       author = {{M{\"u}ller}, Bernhard and {Heger}, Alexander and {Powell}, Jade},
        title = "{Minimum Neutron Star Mass in Neutrino-Driven Supernova Explosions}",
      journal = {\prl},
         year = 2025,
        month = feb,
       volume = {134},
       number = {7},
          eid = {071403},
        pages = {071403},
          doi = {10.1103/PhysRevLett.134.071403},
archivePrefix = {arXiv},
       eprint = {2407.08407},
 primaryClass = {astro-ph.HE},
       adsurl = {https://ui.adsabs.harvard.edu/abs/2025PhRvL.134g1403M}
}

@ARTICLE{WangBurrows2025,
       author = {{Wang}, Tianshu and {Burrows}, Adam},
        title = "{The Effect of the Fast-flavor Instability on Core-collapse Supernova Models}",
      journal = {\apj},
         year = 2025,
        month = jun,
       volume = {986},
       number = {2},
          eid = {153},
        pages = {153},
          doi = {10.3847/1538-4357/add889},
archivePrefix = {arXiv},
       eprint = {2503.04896},
 primaryClass = {astro-ph.HE},
       adsurl = {https://ui.adsabs.harvard.edu/abs/2025ApJ...986..153W}
}

@ARTICLE{SykesMuller2025,
       author = {{Sykes}, Bailey and {M{\"u}ller}, Bernhard},
        title = "{Long-time 3D supernova simulations of nonrotating progenitors with magnetic fields}",
      journal = {\prd},
         year = 2025,
        month = mar,
       volume = {111},
       number = {6},
          eid = {063042},
        pages = {063042},
          doi = {10.1103/PhysRevD.111.063042},
archivePrefix = {arXiv},
       eprint = {2412.01155},
 primaryClass = {astro-ph.HE},
       adsurl = {https://ui.adsabs.harvard.edu/abs/2025PhRvD.111f3042S}
}

@ARTICLE{Janka2025,
       author = {{Janka}, H. -Thomas},
        title = "{Long-Term Multidimensional Models of Core-Collapse Supernovae: Progress and Challenges}",
      journal = {arXiv e-prints},
         year = 2025,
        month = feb,
          eid = {arXiv:2502.14836},
        pages = {arXiv:2502.14836},
          doi = {10.48550/arXiv.2502.14836},
archivePrefix = {arXiv},
       eprint = {2502.14836},
 primaryClass = {astro-ph.HE},
       adsurl = {https://ui.adsabs.harvard.edu/abs/2025arXiv250214836J}
}

@ARTICLE{ParadisoCoughlin2025,
       author = {{Paradiso}, Daniel A. and {Coughlin}, Eric R.},
        title = "{Gotta Go Fast: A Generalization of the Escape Speed to Fluid-dynamical Explosions and Implications for Astrophysical Transients}",
      journal = {\apj},
         year = 2025,
        month = jun,
       volume = {985},
       number = {2},
          eid = {173},
        pages = {173},
          doi = {10.3847/1538-4357/adce6f},
archivePrefix = {arXiv},
       eprint = {2504.11527},
 primaryClass = {astro-ph.HE},
       adsurl = {https://ui.adsabs.harvard.edu/abs/2025ApJ...985..173P}
}

@ARTICLE{Braudoetal2025,
       author = {{Braudo}, Jessica and {Michaelis}, Amir and {Akashi}, Muhammad and {Soker}, Noam},
        title = "{Simulating the Shaping of Point-symmetric Structures in the Jittering Jets Explosion Mechanism}",
      journal = {\pasp},
         year = 2025,
        month = may,
       volume = {137},
       number = {5},
          eid = {054201},
        pages = {054201},
          doi = {10.1088/1538-3873/add08e},
archivePrefix = {arXiv},
       eprint = {2503.10326},
 primaryClass = {astro-ph.HE},
       adsurl = {https://ui.adsabs.harvard.edu/abs/2025PASP..137e4201B}
}

@ARTICLE{Izzoetal2019,
       author = {{Izzo}, L. and {de Ugarte Postigo}, A. and {Maeda}, K. and {Th{\"o}ne}, C.~C. and {Kann}, D.~A. and {Della Valle}, M. and {Sagues Carracedo}, A. and {Micha{\l}owski}, M.~J. and {Schady}, P. and {Schmidl}, S. and {Selsing}, J. and {Starling}, R.~L.~C. and {Suzuki}, A. and {Bensch}, K. and {Bolmer}, J. and {Campana}, S. and {Cano}, Z. and {Covino}, S. and {Fynbo}, J.~P.~U. and {Hartmann}, D.~H. and {Heintz}, K.~E. and {Hjorth}, J. and {Japelj}, J. and {Kami{\'n}ski}, K. and {Kaper}, L. and {Kouveliotou}, C. and {Kru{\.Z}y{\'n}ski}, M. and {Kwiatkowski}, T. and {Leloudas}, G. and {Levan}, A.~J. and {Malesani}, D.~B. and {Micha{\l}owski}, T. and {Piranomonte}, S. and {Pugliese}, G. and {Rossi}, A. and {S{\'a}nchez-Ram{\'\i}rez}, R. and {Schulze}, S. and {Steeghs}, D. and {Tanvir}, N.~R. and {Ulaczyk}, K. and {Vergani}, S.~D. and {Wiersema}, K.},
        title = "{Signatures of a jet cocoon in early spectra of a supernova associated with a {\ensuremath{\gamma}}-ray burst}",
      journal = {\nat},
         year = 2019,
        month = jan,
       volume = {565},
       number = {7739},
        pages = {324-327},
          doi = {10.1038/s41586-018-0826-3},
archivePrefix = {arXiv},
       eprint = {1901.05500},
 primaryClass = {astro-ph.HE},
       adsurl = {https://ui.adsabs.harvard.edu/abs/2019Natur.565..324I}
}

@ARTICLE{Bambaetal2025CasA,
       author = {{Bamba}, Aya and {Agarwal}, Manan and {Vink}, Jacco and {Plucinsky}, Paul and {Terada}, Yukikatsu and {Behar}, Ehud and {Katsuda}, Satoru and {Mori}, Koji and {Sawada}, Makoto and {Matsumoto}, Hironori and {Corrales}, Lia and {Foster}, Adam and {Fujimoto}, Shin-ichiro and {Gu}, Liyi and {Ichikawa}, Kazuhiro and {Matsunaga}, Kai and {Mizuno}, Tsunefumi and {Murakami}, Hiroshi and {Nakajima}, Hiroshi and {Sato}, Toshiki and {Sonoda}, Haruto and {Suzuki}, Shunsuke and {Tateishi}, Dai and {Uchida}, Hiroyuki and {Ichihashi}, Masahiro and {Nobukawa}, Kumiko and {Orlando}, Salvatore},
        title = "{Measuring the asymmetric expansion of the Fe ejecta of Cassiopeia A with XRISM/Resolve}",
      journal = {\pasj},
         year = 2025,
        month = may,
          doi = {10.1093/pasj/psaf041},
archivePrefix = {arXiv},
       eprint = {2504.03268},
 primaryClass = {astro-ph.HE},
       adsurl = {https://ui.adsabs.harvard.edu/abs/2025PASJ..tmp...58B}
}

@ARTICLE{Shishkinetal2025S147,
       author = {{Shishkin}, Dmitry and {Bear}, Ealeal and {Soker}, Noam},
        title = "{Natal kick by early-asymmetrical pairs of jets to the neutron star of supernova remnant S147}",
      journal = {arXiv e-prints},
         year = 2025,
        month = jun,
          eid = {arXiv:2506.21548},
        pages = {arXiv:2506.21548},
archivePrefix = {arXiv},
       eprint = {2506.21548},
 primaryClass = {astro-ph.HE},
       adsurl = {https://ui.adsabs.harvard.edu/abs/2025arXiv250621548S}
}

@ARTICLE{Vinketal2025,
       author = {{Vink}, Jacco and {Agarwal}, Manan and {Bamba}, Aya and {Gu}, Liyi and {Plucinsky}, Paul and {Behar}, Ehud and {Corrales}, Lia and {Foster}, Adam and {Fujimoto}, Shin-ichiro and {Ichihashi}, Masahiro and {Ichikawa}, Kazuhiro and {Katsuda}, Satoru and {Matsumoto}, Hironori and {Matsunaga}, Kai and {Mizuno}, Tsunefumi and {Mori}, Koji and {Murakami}, Hiroshi and {Nakajima}, Hiroshi and {Sato}, Toshiki and {Sawada}, Makoto and {Sonoda}, Haruto and {Suzuki}, Shunsuke and {Tateishi}, Dai and {Terada}, Yukikatsu and {Uchida}, Hiroyuki},
        title = "{Mapping Cassiopeia A's silicon/sulfur Doppler velocities with XRISM-Resolve}",
      journal = {arXiv e-prints},
         year = 2025,
        month = may,
          eid = {arXiv:2505.04691},
        pages = {arXiv:2505.04691},
          doi = {10.48550/arXiv.2505.04691},
archivePrefix = {arXiv},
       eprint = {2505.04691},
 primaryClass = {astro-ph.HE},
       adsurl = {https://ui.adsabs.harvard.edu/abs/2025arXiv250504691V}
}

@ARTICLE{Morietal2025,
       author = {{Mori}, Kanji and {Takiwaki}, Tomoya and {Kotake}, Kei and {Horiuchi}, Shunsaku},
        title = "{Three-dimensional core-collapse supernova models with phenomenological treatment of neutrino flavor conversions}",
      journal = {\pasj},
         year = 2025,
        month = apr,
       volume = {77},
       number = {2},
        pages = {L9-L15},
          doi = {10.1093/pasj/psaf007},
archivePrefix = {arXiv},
       eprint = {2501.15256},
 primaryClass = {astro-ph.HE},
       adsurl = {https://ui.adsabs.harvard.edu/abs/2025PASJ...77L...9M}
}

@ARTICLE{Soker2025Padova,
       author = {{Soker}, Noam},
        title = "{The primary role of jets in powering core-collapse supernovae}",
      journal = {Video Memorie della Societa Astronomica Italiana},
         year = 2025,
        month = apr,
       volume = {2},
          eid = {47},
        pages = {47},
          doi = {10.36116/VIDEOMEM_2.2025.47},
       adsurl = {https://ui.adsabs.harvard.edu/abs/2025VMSAI...2...47S}
}

@ARTICLE{Maltsevetal2025,
       author = {{Maltsev}, K. and {Schneider}, F.~R.~N. and {Mandel}, I. and {M{\"u}ller}, B. and {Heger}, A. and {R{\"o}pke}, F.~K. and {Laplace}, E.},
        title = "{Explodability criteria for the neutrino-driven supernova mechanism}",
      journal = {\aap},
         year = 2025,
        month = aug,
       volume = {700},
          eid = {A20},
        pages = {A20},
          doi = {10.1051/0004-6361/202554931},
archivePrefix = {arXiv},
       eprint = {2503.23856},
 primaryClass = {astro-ph.SR},
       adsurl = {https://ui.adsabs.harvard.edu/abs/2025A&A...700A..20M}
}

@ARTICLE{Orlandoetal20251987A,
       author = {{Orlando}, S. and {Miceli}, M. and {Ono}, M. and {Nagataki}, S. and {Aloy}, M. -A. and {Bocchino}, F. and {Gabler}, M. and {Giudici}, B. and {Giuffrida}, R. and {Greco}, E. and {La Malfa}, G. and {Lee}, S. -H. and {Obergaulinger}, M. and {Petruk}, O. and {Sapienza}, V. and {Ustamujic}, S. and {Weng}, J.},
        title = "{Tracing the ejecta structure of supernova 1987A: Insights and diagnostics from 3D magnetohydrodynamic simulations}",
      journal = {\aap},
         year = 2025,
        month = jul,
       volume = {699},
          eid = {A305},
        pages = {A305},
          doi = {10.1051/0004-6361/202554862},
archivePrefix = {arXiv},
       eprint = {2504.19896},
 primaryClass = {astro-ph.HE},
       adsurl = {https://ui.adsabs.harvard.edu/abs/2025A&A...699A.305O}
}

@ARTICLE{Janka2025Padova,
       author = {{Janka}, Thomas},
        title = "{Core-collapse Supernova Theory in 2025: Progress and Puzzles}",
      journal = {Video Memorie della Societa Astronomica Italiana},
         year = 2025,
        month = apr,
       volume = {2},
          eid = {46},
        pages = {46},
          doi = {10.36116/VIDEOMEM_2.2025.46},
       adsurl = {https://ui.adsabs.harvard.edu/abs/2025VMSAI...2...46J}
}

@ARTICLE{SokerAkashi2025,
       author = {{Soker}, Noam and {Akashi}, Muhammad},
        title = "{The explosion jets of the core-collapse supernova remnant Circinus X-1}",
      journal = {arXiv e-prints},
         year = 2025,
        month = aug,
          eid = {arXiv:2508.10843},
        pages = {arXiv:2508.10843},
          doi = {10.48550/arXiv.2508.10843},
archivePrefix = {arXiv},
       eprint = {2508.10843},
 primaryClass = {astro-ph.HE},
       adsurl = {https://ui.adsabs.harvard.edu/abs/2025arXiv250810843S}
}

@ARTICLE{WangShishkinSoker2025,
       author = {{Wang}, Nikki Yat Ning and {Shishkin}, Dmitry and {Soker}, Noam},
        title = "{Jittering jets in stripped-envelope core-collapse supernovae}",
      journal = {arXiv e-prints},
         year = 2025,
        month = oct,
          eid = {arXiv:2510.02203},
        pages = {arXiv:2510.02203},
          doi = {10.48550/arXiv.2510.02203},
archivePrefix = {arXiv},
       eprint = {2510.02203},
 primaryClass = {astro-ph.HE},
       adsurl = {https://ui.adsabs.harvard.edu/abs/2025arXiv251002203W}
}

@ARTICLE{Soker2025N132D,
       author = {{Soker}, Noam},
        title = "{Attributing the point symmetric structure of core-collapse supernova remnant N132D to the jittering jets explosion mechanism}",
      journal = {arXiv e-prints},
         year = 2025,
        month = jul,
          eid = {arXiv:2507.00757},
        pages = {arXiv:2507.00757},
          doi = {10.48550/arXiv.2507.00757},
archivePrefix = {arXiv},
       eprint = {2507.00757},
 primaryClass = {astro-ph.HE},
       adsurl = {https://ui.adsabs.harvard.edu/abs/2025arXiv250700757S}
}

@ARTICLE{Yamazakietal2014RAA,
       author = {{Yamazaki}, Ryo and {Ohira}, Yutaka and {Sawada}, Makoto and {Bamba}, Aya},
        title = "{Synchrotron X-ray diagnostics of cutoff shape of nonthermal electron spectrum at young supernova remnants}",
      journal = {Research in Astronomy and Astrophysics},
         year = 2014,
        month = feb,
       volume = {14},
       number = {2},
          eid = {165-178},
        pages = {165-178},
          doi = {10.1088/1674-4527/14/2/005},
archivePrefix = {arXiv},
       eprint = {1301.7499},
 primaryClass = {astro-ph.HE},
       adsurl = {https://ui.adsabs.harvard.edu/abs/2014RAA....14..165Y}
}
\bibliographystyle{aasjournal}
  


\end{document}